\documentclass[runningheads]{rnl}

\usepackage[utf8]{inputenc}      
\usepackage[english]{babel}

\usepackage{graphicx, subfigure}  

\usepackage{amsmath,amsfonts,amssymb,epstopdf}   

\usepackage[sectionbib,super]{natbib}
\usepackage{indentfirst}

\begin{document}
 
\title*{Diffusionless hydromagnetic modes in rotating ellipsoids: a road to weakly nonlinear models?}
 
 
\titrecourt{Modes of rotating ellipsoids}

\author{Jérémie Vidal
\and David Cébron
\and Nathanaël Schaeffer}

\index{Vidal Jérémie}              
\index{Cébron David}
\index{Schaeffer Nathanaël}

\auteurcourt{Vidal {\it et al.}}
 
\adresse{Universit\'e Grenoble Alpes, CNRS, ISTerre, Grenoble, France.}

\email{jeremie.vidal@univ-grenoble-alpes.fr}

\maketitle

\begin{resume}
On étudie les modes hydromagnétiques d'un fluide incompressible contenu dans un ellipsoide triaxial quelconque en rotation. Le vecteur rotation est incliné par rapport aux axes principaux de l'ellipsoide. L'état de base est constitué d'une densité de courant uniforme dans les coordonnées d'espace et inclinée par rapport aux axes d'inertie.
On projette les perturbations tri-dimensionnelles sur un espace vectoriel de dimension finie. En combinant calcul symbolique et numérique, nous pouvons calculer les modes propres avec des compléxités spatiales élevées. Les résultats précédemment obtenus dans la sphère sont généralisés à la géométrie triaxiale et comparés à une analyse locale en ondes planes. En l'absence de champ magnétique, les modes se réduisent aux modes inertiels de l'ellipsoide triaxial, qui forment une base complète. Nous utiliserons ces modes pour étudier la saturation faiblement non-linéaire des instabilities inertielles, dont l'instabilité elliptique.
\end{resume}

\begin{resumanglais}
We investigate free hydromagnetic eigenmodes of an incompressible, inviscid and ideal electrically conducting fluid in rotating triaxial ellipsoids.
The container rotates with an angular velocity tilted from its figure. The magnetic base state is a uniform current density also tilted. Three-dimensional perturbations upon the base state are expanded onto a finite-dimensional polynomial basis. By combining symbolic and numeric computations, we are able to get the eigenmodes of high spatial complexity. Hydromagnetic modes of the sphere still exist in triaxial geometry. A plane-wave analysis is also carried on, explaining the dispersion relation observed in our model. Without magnetic field, the modes reduce to the inertial modes of the ellipsoids, which form a complete basis. We propose to use these modes to study the weakly non-linear saturation of inertial instabilities, especially the elliptical one.
\end{resumanglais}

\section{Introduction}
Hydromagnetic modes are thought to play an important role in the dynamics of magnetized liquid metal planetary cores \citep{vidalref_gillet2010fast, vidalref_labbe2015magnetostrophic, vidalref_le2013waves, vidalref_malkus1967hydromagnetic}.
These waves are influenced by both the fast rotation of the core and the magnetic field that permeates the fluid. Some theoretical studies have focused on hydromagnetic modes growing upon a toroidal magnetic field in spherical geometry, motivated by the existence of a strong toroidal magnetic field in the Earth's core. \citet{vidalref_malkus1967hydromagnetic} shows that, for a well chosen axisymmetric and azimuthal toroidal field (the Malkus field), the hydromagnetic oscillations in spherical containers are governed by a modified version of the Poincar\'e equation (the governing equation of inertial modes whose restoring force is the Coriolis force).
This observation enables Malkus to use the properties of the Poincar\'e equation to determine dispersion relation of the associated hydromagnetic modes. Based on the pioneering work of Malkus, \citet{vidalref_zhang2003nonaxisymmetric} give the explicit eigenvalues, thanks to the solutions of the original Poincar\'e equation \citep{vidalref_zhang2001inertial}. With another approach, \citet{vidalref_labbe2015magnetostrophic} consider non axisymmetric background fields, improving our understanding of waves in spherical geometry. Kerswell 
\citep{vidalref_kerswell1993elliptical,vidalref_kerswell1994tidal} considers a spheroidal container, modelling the tidal deformation of the Earth's core due to the gravitational torques exerted by the Sun and the Moon.
He also find the dispersion relation of the hydromagnetic modes by modifying the modal frequencies of pure inertial modes.
However this approach cannot be used in triaxial geometry, because pure inertial modes are not known analytically.
Instead, \citet{vidalref_vantieghem2014inertial} proposes a numerical algorithm to compute them in triaxial ellipsoids, but the study is restricted to modes with linear and quadratic dependence on space coordinates.

We extend the theory of free hydromagnetic modes to triaxial geometries, applying the method of \citet{vidalref_vantieghem2014inertial} to the hydromagnetic case.
Our study is restricted to background magnetic fields which are toroidal and depend linearly on space coordinates.
This choice enables to take into account non-axisymmetric fields in ellipsoidal geometry.
Despite the strong assumptions, this model is a reasonable starting point to study free hydromagnetic eigenmodes in planetary cores.
Finally, we briefly discuss the usefulness of these modes to build weakly nonlinear models of any inertial instability in rotating ellipsoids, in particular the elliptical instability.

\section{Tilted hydromagnetic eigenmodes}
	\subsection{Governing equations}
\begin{figure}[t]
	\centering
	\includegraphics[width=0.8\textwidth]{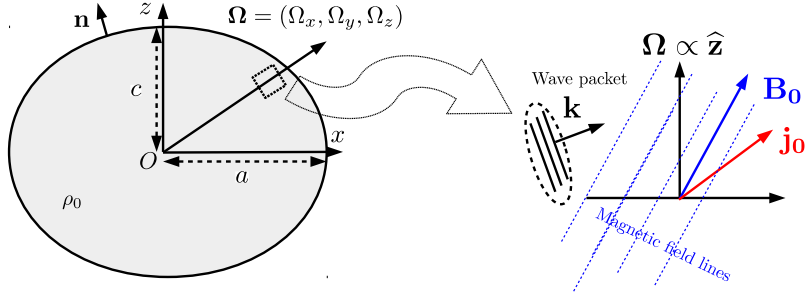}
	\caption{(Left) Ellipsoidal geometry in the plane ($Ox,Oz$). $\mathbf{n}$ is the unitary vector normal to the ellipsoidal boundary. (Right) Infinite plane layer setup for hydromagnetic waves. The imposed magnetic field $\mathbf{B_0}$ and current density $\mathbf{j_0}$ are uniform in the local analysis. $\mathbf{k}$ is the local wave number.}
	\label{vidalfig_Fig_Ellipsoid}
\end{figure}

We consider a homogeneous, incompressible, inviscid and ideal electrically conducting fluid enclosed in a rigid triaxial ellipsoid, rotating at the constant angular velocity $\boldsymbol{\Omega}$. The triaxial shape results from gravitational forces exerted by an orbital partner (moon, star\dots).
We work in the reference frame attached with the principal axis of length $a,b,c$.
The angular velocity $\boldsymbol{\Omega} = (\Omega_x, \Omega_y, \Omega_z)$ is tilted from the figure axis, as illustrated in Figure \ref{vidalfig_Fig_Ellipsoid} (Left).
The fluid is permeated by an imposed magnetic field $\mathbf{B_0}$.
Finally, we do not consider any background velocity field ($\mathbf{U_0} = \mathbf{0}$).
Let us take $|\boldsymbol{\Omega}|^{-1}$ as time scale, a typical length $L$ as length scale, $|\boldsymbol{\Omega}| L$ as velocity scale and the strength of the background magnetic field $B_0$ as magnetic scale.
The linearized dimensionless equations are
\begin{align}
	\lambda \, \mathbf{u} + 2 \, \boldsymbol{\Omega} \, \times \mathbf{u} &= - \nabla \pi + Le^2 \left [ (\nabla \times \mathbf{B_0}) \times \mathbf{b} + (\nabla \times \mathbf{b}) \times \mathbf{B_0} \right ],\label{vidaleq_Eq_Euler_Mode} \\
	\lambda \, \mathbf{b} &= \nabla \times (\mathbf{u} \times \mathbf{B_0}), \label{vidaleq_Eq_Induction_Mode} \\
	\nabla \cdot \mathbf{u} &= \nabla \cdot \mathbf{b} = 0,
\end{align}
where ($\mathbf{u},\mathbf{b}$) are the velocity and magnetic perturbations, $\lambda = \sigma + \mathrm{i} \omega$ the eigenvalue with $\sigma$ the damping (or growth) rate and $\omega$ the frequency, $\pi$ the reduced pressure and the Lehnert number
\begin{equation}
	Le = \frac{B_0}{\sqrt{\rho_0 \mu_0} \, |\boldsymbol{\Omega}| L},
\end{equation}
with $\rho_0$ the fluid density and $\mu_0$ magnetic permeability of free space.
The dimensionless number $Le$ measures the magnetic field compared to rotation effects.

The velocity field satisfies the impermeability boundary condition $\mathbf{u} \cdot \mathbf{n} = 0$ at the ellipsoidal boundary, where $\mathbf{n}$ is the outward unitary vector normal to the boundary.
The boundary condition on $\mathbf{b}$ is constrained by the one on $\mathbf{u}$ through the equation (\ref{vidaleq_Eq_Induction_Mode}).
Let us denote $\nabla_\mathcal{S} = \nabla - \mathbf{n} \, \partial_n$ , $(B_n, b_n, u_n)$ the normal components of $(\mathbf{B}, \mathbf{b}, \mathbf{u})$ and $\mathbf{u_\mathcal{S}}$ the tangential velocity at the boundary.
Following \citet{vidalref_backus1996foundations}, the scalar product of (\ref{vidaleq_Eq_Induction_Mode}) with  $\mathbf{n}$ leads to the boundary condition satisfied by the normal component $b_n$
\begin{equation}
	\lambda \, b_n = - \nabla_\mathcal{S} (B_n \, \mathbf{u_\mathcal{S}}).
	\label{vidaleq_Eq_BC_bn}
\end{equation}

	\subsection{Background magnetic field $\mathbf{B_0}$}
We consider only background fields such that the associated density currents $\mathbf{j_0} = \nabla \times \mathbf{B_0}$ are uniform in space. The background field $\mathbf{B_0}$ is thus a linear combination of the following  elements
\begin{equation}
	\mathbf{B_{01}} = \left ( 0, - \frac{z}{c^2}, \frac{y}{b^2} \right ), \ \mathbf{B_{02}} = \left ( \frac{z}{c^2}, 0, -\frac{x}{a^2} \right ), \ \mathbf{B_{03}} = \left ( -\frac{y}{b^2}, \frac{x}{a^2}, 0 \right ). \label{Eq_B013} 
\end{equation}
In spherical geometry, $\mathbf{B_{03}}$ is the Malkus field \citep{vidalref_malkus1967hydromagnetic}, an axisymmetric toroidal field which is simply given by the cylindrical radius. Here our general background field is not axisymmetric, allowing the coupling of modes of different azimuthal wave numbers, even in spherical or spheroidal geometry. Fields (\ref{Eq_B013}) are purely toroidal, i.e. $\mathbf{B_0} \cdot \mathbf{n} = 0$ everywhere.
Thus the boundary condition (\ref{vidaleq_Eq_BC_bn}) for the magnetic perturbation reduces to $\mathbf{b} \cdot \mathbf{n} = 0$ at the outer boundary.

	\subsection{Global polynomial basis}
We considers the finite-dimensional vector space $\boldsymbol{\mathcal{V}}_{n}$ of polynomial flows of maximum degree $n$ which are solenoidal and satisfy the impermeability boundary condition in triaxial ellipsoids.
As in the spherical case \citep{vidalref_ivers2015enumeration}, the dimension of $\boldsymbol{\mathcal{V}}_{n} $ is $\mathcal{D}_n = n (n+1) (2n+7)/6$. $\boldsymbol{\mathcal{V}}_{n}$ is an invariant of equations (\ref{vidaleq_Eq_Euler_Mode}) - (\ref{vidaleq_Eq_Induction_Mode}). Following \citet{vidalref_wu2011high}, we build an explicit polynomial basis of $\boldsymbol{\mathcal{V}}_{n}$ for any degree $n$, on which both $\mathbf{u}$ and $\mathbf{b}$ are expanded. This leads to a generalised eigenvalue problem of size $4 \mathcal{D}_n^2$, where the eigenvectors are the $2 \mathcal{D}_n$ projection coefficients of the linear combination of the basis elements for a given degree $n$.

\section{Results}
	\subsection{Local analysis of hydromagnetic waves}
\begin{figure}[t]
	\centering
	\includegraphics[width=0.95\textwidth]{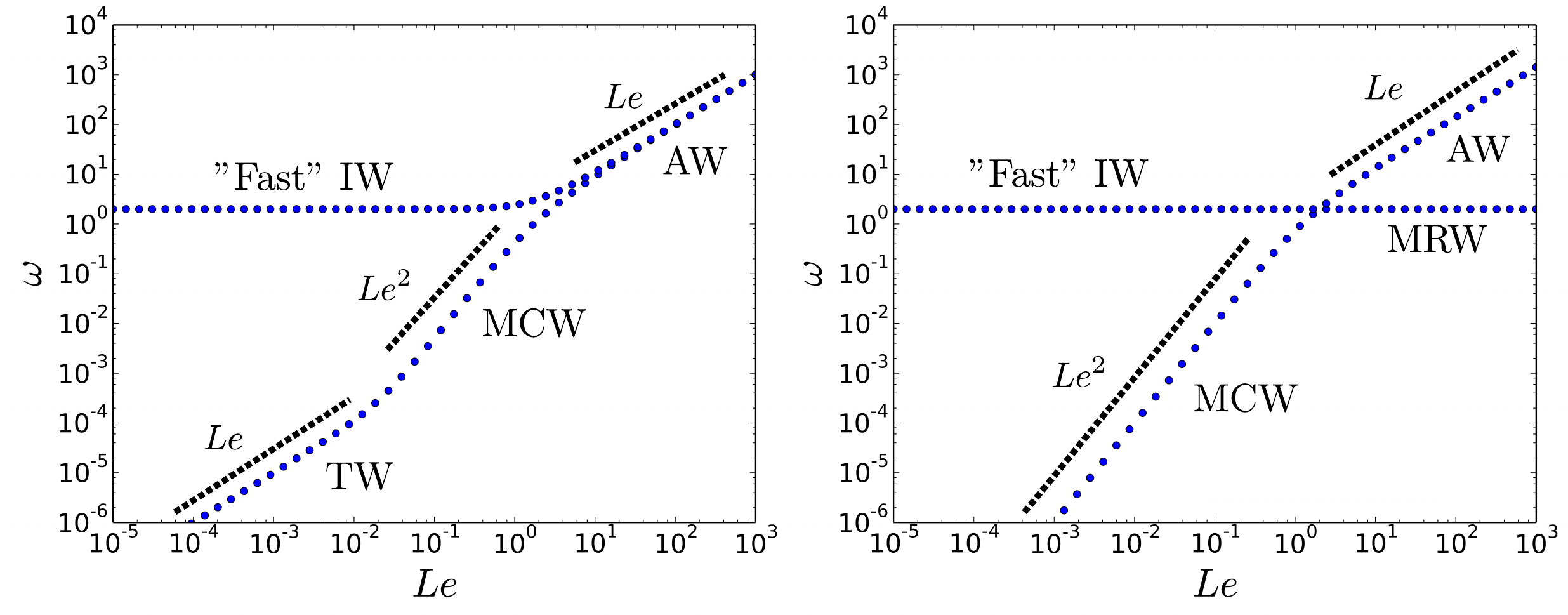}
	\caption{Dispersion curves obtained with the local plane wave analysis. AW: Alfv\'en Waves, IW: Inertial Waves, MCW: Magneto-Coriolis Waves, MRM: "Magneto-Rotational" Waves, TW: Torsional Waves. (Left) Case 1: $\mathbf{j_0} = \mathbf{0}$, $\mathbf{B_0} = (0.01,0,1)$ and $\mathbf{k} = (0,0,1)$. TW and MCW branches cross for $Le \lesssim 2 B_x/(k_z B_z^2)$, whereas IW and AW branches cross at $Le \simeq 1$. (Right) Case 2: $\mathbf{j_0} = (0,0,1)$, $\mathbf{B_0} = (0,0,1)$ and $\mathbf{k} = (0,0,1)$.}
	\label{vidalfig_Fig_Freq_Plane_Wave}
\end{figure}

We can infer some properties of the waves propagating in the system with a simple plane-wave analysis in Cartesian coordinates, extending the one of \citet{vidalref_galtier2014weak}. As shown in Figure \ref{vidalfig_Fig_Ellipsoid} (Right), we consider a dimensionless rotation vector $\boldsymbol{\Omega} = (0,0,1)$ aligned  with the $\mathbf{\widehat{z}}$-axis. We assume that both the background magnetic field $\mathbf{B_0}$ and current density  $\mathbf{j_0}$ are uniform and may be inclined from the spin axis. Without loss of generality, we study two simple configurations such that analytic expressions of the dispersion relation are available. 

We first consider $\mathbf{j_0}=\mathbf{0}, \mathbf{B_0} = (B_x,0,B_z)$ and the wave number $\mathbf{k} = (0,0,k_z)$. The dispersion relation
\begin{equation}
	\lambda = \pm \frac{1}{2} \sqrt{\pm 2 \sqrt{16 +16 B_z^2 Le^2 k_z^2 -8 B_x^2 Le^2 k_z^2+B_x^4 Le^4 k_z^4} - 8 - 4 B_z^2 Le^2 k_z^2 - 2 B_x^2 Le^2 k_z^2},
	\label{vidaleq_Eq_DC_Case1}
\end{equation}
illustrated in Figure \ref{vidalfig_Fig_Freq_Plane_Wave} (Left), allows to recover the usual waves frequencies. Indeed, for $Le \gg 1$ it gives $\lambda \simeq \pm \mathrm{i} k_z B_z Le$ and $\lambda \simeq \pm \mathrm{i} k_z Le \sqrt{B_x^2+B_z^2} $, i.e. the "fast" Alfv\'en waves (AW) weakly affected by the rotation. For $ Le \ll 1$ equation (\ref{vidaleq_Eq_DC_Case1}) gives $\lambda \simeq \pm 2 \mathrm{i}  $, which are the "fast" inertial waves (IW) weakly affected by $\mathbf{B_0}$, but also  
\begin{equation}
	\lambda \simeq \pm \mathrm{i} k_z Le \sqrt{B_x^2+B_z^4 k_z^2 Le^2/4}.
	\label{vidaleq_Eq_DC_Case11}
\end{equation}
When $B_x^2 \ll B_z^4 k_z^2 Le^2/4$, equation (\ref{vidaleq_Eq_DC_Case11}) gives $\lambda \simeq \pm \mathrm{i}  k_z^2 B_z^2 Le^2/2$, which corresponds to the "slow" magneto-Coriolis waves (MCW) branch. These waves are affected by both rotation and magnetic effects. When $B_x^2 \gg B_z^4 k_z^2 Le^2/4$ equation (\ref{vidaleq_Eq_DC_Case11}) gives $\lambda \simeq \pm \mathrm{i} k_z B_x Le$, which corresponds to the torsional waves (TW) branch. The TW, appearing at $Le \lesssim 2 B_x/(k_z B_z^2)$, are "slow" Alfv\'en waves emerging from the MCW branch, because the spin axis $\boldsymbol{\Omega}$ is not aligned with the background field $\mathbf{B_0}$.

\begin{figure}[t]
	\centering
	\includegraphics[width=0.95\textwidth]{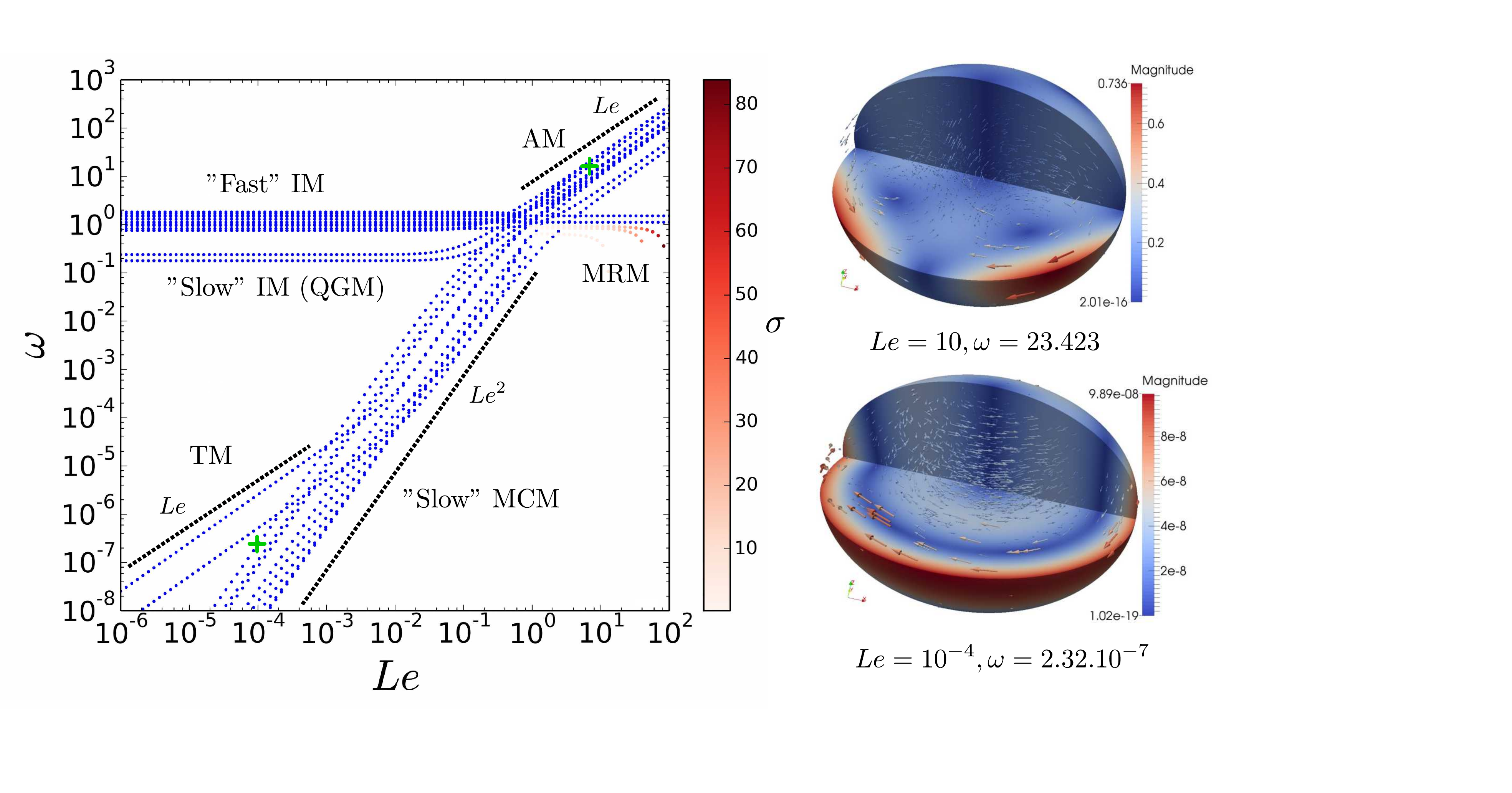}
	\caption{Effect of $Le$ on the dimensionless frequency $\omega$ of hydromagnetic modes in an ellipsoid ($a=1$, $b=0.9$ and $c=0.7$).
	The background magnetic field is $0.01 \, \mathbf{B_{01}} + \mathbf{B_{03}}$. The rotation vector is aligned with the $c$ axis and has a magnitude $|\boldsymbol{\Omega}| = 1$.
	For the sake of clarity, only modes up to $n=3$ are shown.
	(Left) Dispersion curve. Each point is a mode. The blue ones are stable ($\sigma \leq 10^{-8}$). The red modes are unstable ($\sigma \geq 10^{-8}$) and their growth rate $\sigma$ is given with the colorbar.
	TM: Torsional modes, IM: Inertial modes, QGM: Quasi-geostrophic modes, MCM: Magneto-Coriolis modes, AM: Alfv\'en modes, MRM: "Magneto-rotational" modes.
	(Right) Isocontours of the velocity magnitude $||\mathbf{u}||$ and velocity vector arrows in two orthogonal planes for two particular modes, represented with green stars in the left figure.}
	\label{vidalfig_Fig_Hydromag}
\end{figure}

We now introduce a background current density and remove the horizontal magnetic field, assuming $\boldsymbol{\Omega} = (0,0,1)$, $\mathbf{j_0} = (0,0,j_z)$, $\mathbf{B_0} = (0,0,B_z)$ and $\mathbf{k} = (0,0,k_z)$.
Even though this choice is not consistent for the basic state, the model contains all the ingredients to  explain the physics of the various hydromagnetic waves, including the effect of an electric current.
The dispersion relation, illustrated in Figure \ref{vidalfig_Fig_Freq_Plane_Wave} (Right), is now
\begin{equation}
	\lambda = \pm \mathrm{i} \pm \mathrm{i} \sqrt{1 + Le^2 B_z k_z ( B_z k_z \pm j_z)}.
	\label{vidaleq_Eq_DC_Case2}
\end{equation}
As explained above, the TW branch disappears because $\mathbf{B_0}$ is here aligned with the spin axis $\boldsymbol{\Omega}$. However IW, MCW and AW branches still exist, with respectively $\lambda \simeq \pm 2 \mathrm{i} $ and $\lambda \simeq  \mathrm{i}  k_z^2 B_z^2 Le^2 [j_z/(k_z B_z) \pm 1]/2$ for $Le \ll 1$, whereas $\lambda \simeq \pm \mathrm{i} \pm \mathrm{i} k_z B_z Le \sqrt{1 \pm j_z/(k_z B_z)} $ for  $Le \gg 1$ (in agreement with previously obtained results). This latter expression shows that another kind of wave appears because of the current density, when $j_z \simeq k_z B_z$, leading to $\lambda \simeq \pm \mathrm{i}$. We call them "magneto-rotational" waves (MRW), as they disappear without rotation. Their frequencies are mainly independent of $Le$, as for the IW branch.

	\subsection{Global modes}
We have benchmarked our numerical results against the explicit modal solutions in spherical geometry for the imposed Malkus field \citep{vidalref_malkus1967hydromagnetic,vidalref_zhang2003nonaxisymmetric}, showing no significant discrepancies between the two approaches.
We sum up the results in triaxial geometry with the Figure \ref{vidalfig_Fig_Hydromag}. For the sake of clarity, only modes of polynomial degrees up to $n=3$ are represented. The background density current is not aligned with the rotation vector, by adding to the Malkus field $\mathbf{B_{03}}$ the non-axisymmetric component $\mathbf{B_{01}}$ with a small amplitude. In the triaxial geometry, all the branches predicted by the local dispersion relations (\ref{vidaleq_Eq_DC_Case1}) - (\ref{vidaleq_Eq_DC_Case2}) are present: the inertial modes (IM), the "slow" magneto-Coriolis modes (MCM), the "fast" Alfv\'en modes (AM), the "slow" torsional modes (TM) and the "magneto-rotational" modes (MRM). It is worth noting that the TM branch does not exist for the pure Malkus field \citep{vidalref_labbe2015magnetostrophic}. We find that the IW, MCM, AM and TW are stable, whereas the MRM branch at $Le > 1$ may be stable or unstable, as predicted by \citet{vidalref_malkus1967hydromagnetic} in spherical geometry. This instability is not relevant for geophysical systems (where $Le \ll 1)$. Finally, the IM branch splits into the "fast" inertial modes and the "slow" quasi-geostrophic inertial modes (QGM). The latter have velocity fields more or less aligned with the spin axis $\boldsymbol{\Omega}$ of the container, reminiscent of the Taylor-Proudman constraint.

\section{Weakly nonlinear models using inertial modes?}
\begin{figure}[t]
	\centering
	\includegraphics[width=0.9\textwidth]{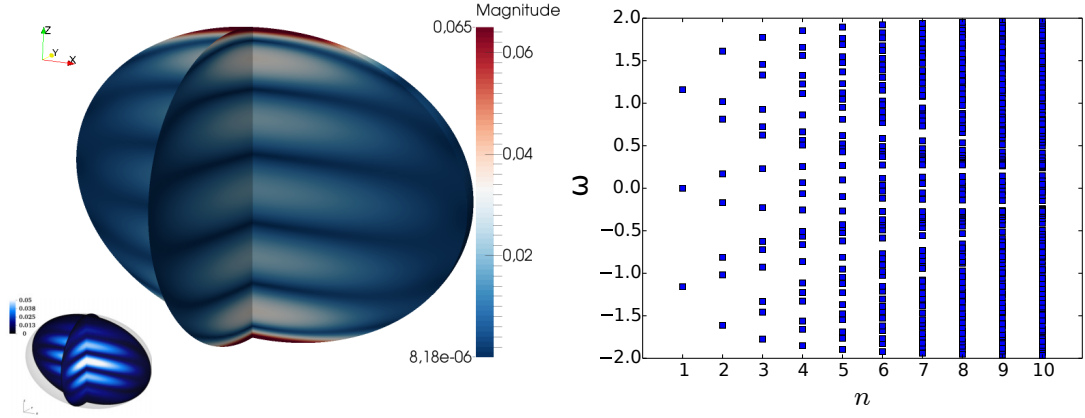}
	\caption{Inertial modes of the ellipsoidal container used by \citet{vidalref_grannan2014experimental,vidalref_favier2015generation} ($a=1, b=0.7, c=0.7$). (Left) Velocity amplitude of an inviscid inertial mode ($n=6$, $\omega = 1.9219$) in two perpendicular planes. It corresponds to the red star in the right figure. The viscous mode, observed in \citet{vidalref_favier2015generation} and filtered at $\omega = 1.95$, is shown in the bottom left corner.  (Right) Dimensionless frequencies $\omega$ in function of the degree $n$.}
	\label{vidalfig_Fig_IM}
\end{figure}

The case of pure inertial modes is also interesting in its own, since they are essential to understand the dynamics of rapidly rotating fluids in geo- and astrophysics. Using the mathematical apparatus developed by \citet{vidalref_ivers2015enumeration}, it can be proved that inertial modes form a complete basis.
The completeness opens exciting perspectives to analyse many geophysical and astrophysical fluid problems, because the rotational effect does not couple the inertial modes. Therefore they may offer an efficient basis to understand bounded rotating fluids in ellipsoids, such as any inertial instability driven by mechanical forcings (tides, precession, libration\dots)

In particular, the elliptical instability growing upon the laminar libration-driven base flow is observed both in the numerical\citep{vidalref_favier2015generation} and laboratory experiments\citep{vidalref_grannan2014experimental}. Its underlying mechanism is the triadic resonance between two inertial modes and the base flow \citep{vidalref_vantieghem2015latitudinal}.
Some viscous inertial modes have been directly observed in the simulations, showing a very good agreement with inviscid modes of the present study except in the viscous boundary layer (Figure \ref{vidalfig_Fig_IM}, Left).
However, the saturation of the instability is not understood. 
Thanks to our tool, building a weakly nonlinear model of the elliptical instability is at reach.
By projecting numerical simulations on a large subset of modes (Figure \ref{vidalfig_Fig_IM}, Right), we could identify the physical modes which are essential for the dynamics.
Then we hope to give a quantitative explanation of the saturation with a low-dimensional weakly nonlinear model of the instability, valid for both laboratory and numerical experiments.

\section{Conclusion and perspectives}
We have revisited the theory of free diffusionless hydromagnetic modes, by considering for the first time triaxial ellipsoidal containers and background magnetic fields of uniform current densities.
The velocity and magnetic perturbations are projected onto a finite-dimensional vector space, made of Cartesian polynomial basis elements satisfying the appropriate boundary conditions.
A code has been developed to solve numerically the symbolic eigenvalue problem.
As a benchmark, it has been applied to the toroidal Malkus field \citep{vidalref_malkus1967hydromagnetic}, for which the mode frequencies are known analytically \citep{vidalref_zhang2003nonaxisymmetric}, showing a very good agreement. The results in triaxial geometry extend the ones of \citet{vidalref_labbe2015magnetostrophic} in spherical geometry.

Note that the chosen base state may not be consistent, as there is no background velocity field sustaining the imposed magnetic field.
Following the approach of Kerswell \citep{vidalref_kerswell1993elliptical,vidalref_kerswell1994tidal}, we have found a consistent and steady magnetostrophic regime in triaxial geometry, assuming a background velocity field of uniform vorticity.
This magnetostrophic regime is a magnetohydrodynamic analogue of hydrodynamic laminar flows of uniform vorticity, studied previously \citep{vidalref_poincare1910precession, vidalref_kerswell1993instability, vidalref_wu2011high}.
The global stability analysis in the magnetostrophic regime will be performed later. 

Finally, the inertial modes can be used to build weakly nonlinear models of any inertial instability in rotating ellipsoids. We shall first use them to understand the saturation of the elliptical instability, well observed in numerical and laboratory experiments. This work is theoretically possible thanks to the completeness of inertial modes in triaxial geometry and will be carried on soon.


{\small

\begin{thebibliography}{}

\bibitem[Backus et~al., 1996]{vidalref_backus1996foundations}
Backus, G., Parker, R.~L., and Constable, C. (1996).
\newblock Cambridge University Press.

\bibitem[Favier et~al., 2015]{vidalref_favier2015generation}
Favier, B., Grannan, A., Le~Bars, M., and Aurnou, J. (2015).
\newblock {\em Physics of Fluids (1994-present)}, 27(6):066601.

\bibitem[Galtier, 2014]{vidalref_galtier2014weak}
Galtier, S. (2014).
\newblock {\em Journal of Fluid Mechanics}, 757:114--154.

\bibitem[Gillet et~al., 2010]{vidalref_gillet2010fast}
Gillet, N., Jault, D., Canet, E., and Fournier, A. (2010).
\newblock {\em Nature}, 465:74--77.

\bibitem[Grannan et~al., 2014]{vidalref_grannan2014experimental}
Grannan, A., Le~Bars, M., C{\'e}bron, D., and Aurnou, J. (2014).
\newblock {\em Physics of Fluids (1994-present)}, 26(12):126601.

\bibitem[Ivers et~al., 2015]{vidalref_ivers2015enumeration}
Ivers, D., Jackson, A., and Winch, D. (2015).
\newblock {\em Journal of Fluid Mechanics}, 766:468--498.

\bibitem[Kerswell, 1993a]{vidalref_kerswell1993elliptical}
Kerswell, R. (1993a).
\newblock {\em Geophysical \& Astrophysical Fluid Dynamics}, 71(1-4):105--143.

\bibitem[Kerswell, 1993b]{vidalref_kerswell1993instability}
Kerswell, R. (1993b).
\newblock {\em Geophysical \& Astrophysical Fluid Dynamics}, 72(1-4):107--144.

\bibitem[Kerswell, 1994]{vidalref_kerswell1994tidal}
Kerswell, R. (1994).
\newblock {\em Journal of Fluid Mechanics}, 274:219--241.

\bibitem[Kerswell and Malkus, 1998]{vidalref_kerswell1998tidal}
Kerswell, R.~R. and Malkus, W.~V. (1998).
\newblock {\em Geophysical Research Letters}, 25(5):603--606.

\bibitem[Labb{\'e} et~al., 2015]{vidalref_labbe2015magnetostrophic}
Labb{\'e}, F., Jault, D., and Gillet, N. (2015).
\newblock {\em Geophysical \& Astrophysical Fluid Dynamics}, 109(6):587--610.

\bibitem[Le Gal., 2013]{vidalref_le2013waves}
Le Gal, P. (2013).
\newblock Springer.

\bibitem[Malkus, 1967]{vidalref_malkus1967hydromagnetic}
Malkus, W.~V. (1967).
\newblock Hydromagnetic planetary waves.
\newblock {\em Journal of Fluid Mechanics}, 28(04):793--802.

\bibitem[Poincar{\'e}, 1910]{vidalref_poincare1910precession}
Poincar{\'e}, H. (1910).
\newblock {\em Bulletin Astronomique, Serie I}, 27:321--356.

\bibitem[Vantieghem, 2014]{vidalref_vantieghem2014inertial}
Vantieghem, S. (2014).
\newblock {\em Proc. Roy. Soc. A}, 470(2168):20140093.

\bibitem[Vantieghem et~al., 2015]{vidalref_vantieghem2015latitudinal}
Vantieghem, S., C{\'e}bron, D., and Noir, J. (2015).
\newblock {\em Journal of Fluid Mechanics}, 771:193--228.

\bibitem[Wu and Roberts, 2011]{vidalref_wu2011high}
Wu, C.-C. and Roberts, P.~H. (2011).
\newblock {\em Geophysical \& Astrophysical Fluid Dynamics}, 105(2-3):287--303.

\bibitem[Zhang et~al., 2001]{vidalref_zhang2001inertial}
Zhang, K., Earnshaw, P., Liao, X., and Busse, F. (2001).
\newblock {\em Journal of Fluid Mechanics}, 437:103--119.

\bibitem[Zhang et~al., 2003]{vidalref_zhang2003nonaxisymmetric}
Zhang, K., Liao, X., and Schubert, G. (2003).
\newblock {\em The Astrophysical Journal}, 585(2):1124.



\end{thebibliography}
}
\end{document}